\documentclass[
    ,final            
  ]
  {aipproc}

\layoutstyle{8x11double}

\begin{document}

\title{SPINSTARS at low metallicities}


\author{G. Meynet}{
  address={Geneva Observatory, University of Geneva, CH-1290 Sauverny, Switzerland}
}

\author{S. Ekstr\"om}{
  address={Geneva Observatory, University of Geneva, CH-1290 Sauverny, Switzerland}
}

\author{A. Maeder}{
  address={Geneva Observatory, University of Geneva, CH-1290 Sauverny, Switzerland}
}

\author{R. Hirschi}{
  address={Keele University, Staffordshire, ST5 5BG, United Kingdom}
}

\author{C. Chiappini}{
  address={Geneva Observatory, University of Geneva, CH-1290 Sauverny, Switzerland}
  ,altaddress={I.N.A.F. Osservatorio Astronomico di Trieste, via G.B. Tiepolo 11, 34131 Trieste, Italy}
}

\author{C. Georgy}{
  address={Geneva Observatory, University of Geneva, CH-1290 Sauverny, Switzerland}
}

\begin{abstract}
The main effect of axial rotation on the evolution of massive PopIII stars is to trigger internal mixing processes which allow stars to produce significant amounts of primary nitrogen 14 and carbon 13. Very metal poor massive stars produce much more primary nitrogen 
than PopIII stars for a given initial mass and rotation velocity. The very metal poor stars undergo strong mass loss induced by rotation. One can distinguish two types of rotationnaly enhanced stellar winds:
1) Rotationally mechanical winds occurs when the
surface velocity reaches the critical velocity at the equator, {\it i.e.} the velocity at which the centrifugal
acceleration is equal to the gravity; 2) Rotationally radiatively line driven winds are a consequence of strong internal mixing which brings large amounts of CNO elements at the surface. This
enhances the opacity and may trigger strong line driven winds. These effects are important for an initial
value of $\upsilon/\upsilon_{\rm crit}$ of 0.54 for a 60 M$_\odot$ at $Z=10^{-8}$, {\it i.e.} for initial values
of $\upsilon/\upsilon_{\rm crit}$ higher than the one ($\sim$0.4) corresponding to observations at solar $Z$. These two effects, strong
internal mixing leading to the synthesis of large amounts of primary nitrogen and important mass losses
induced by rotation, occur for $Z$ between about 10$^{-8}$ and 0.001.
For metallicities above 0.001 and for reasonable choice of the rotation velocities, internal mixing
is no longer efficient enough to trigger these effects. 

\end{abstract}

\maketitle

\section{Effects of rotation}

Rotation deforms the stars, triggers instabilities as shear turbulence and meridional currents that drive transport of chemical species (called hereafter rotational mixing) and of angular momentum and has an impact on the way massive stars are losing mass (see for more details the review \cite{MMARAA2000}). 

The efficiency of rotational mixing (measured for instance by the degree of surface enrichments at a given
evolutionary stage) increases when the initial mass and rotation increase \cite{MMV}. Very interestingly, this efficiency increases also when the initial metallicity {\it decreases} \cite{MMVII}. Physically, this
is due to the fact that when the metallicty is lower, the stars are more compact. This makes the gradients
of the angular velocity steeper in the stellar interiors. Steeper gradients produce stronger shear turbulence and thus more mixing.   

Non-rotating PopIII and very metal poor stars are expected to lose very small amounts of mass through line driven winds \cite{Kud2002}. For rotating models new effects enter into play which tend to increase the mass lost through stellar winds.
The (line driven) mass loss of a hot massive star at a given position in the HR diagram and
and for a fixed surface abundance is enhanced
by rotation \cite{MMVI} by at most a factor $\sim 2$. This is true provided the luminosity of the star
is well below the Eddington limit ($L < 0.639L_{\rm edd}$). Near the Eddington limit, mass loss might be
increased by much greater factors.

Rotational mixing can bring to the surface heavy elements newly synthesized in the stellar core. Rotation thus produces an increase of the opacity of the outer layers and activates strong mass loss through radiatively line driven winds. This effect may be responsible for the loss of large fractions of the initial mass of the star
(\cite{Meynetal06}; \cite{Hir07}).

Rotation induces wind anisotropies (\cite{Owocki96}; \cite{Maed99}). In the hot part of the HR diagram, a fast rotating
massive star is expected to have polar winds. This process allows the star to lose large amounts of material without losing a lot of angular momentum and thus may be of first importance in the evolutionary scenarios leading to Gamma Ray Bursts \cite{MMGRB}.  

Stars rotating at the critical velocity launch matter into a keplerian equatorial disk.
The disk is probably destroyed by the pressure exerted by the stellar radiation and finally matter is lost.
Such a process seems to occur around Be stars which are stars rotating at or very near the critical limit (see the review by \cite{PorRiv2003}). Hereafter we call ``mechanical winds'' such a process.

Finally rotation modifies the evolutionary track in the HR diagram, making them bluer and more luminous
\cite{MMV}. This of course indirectly affects the way stars are losing mass. Also rotational mixing favors the entry of the stars into the Wolf-Rayet phase during which strong mass loss uncover the stellar cores \cite{MMX}.

Let us recall that rotation improves a lot the stellar models. It allows to reproduce the observed changes of the surface abundances (\cite{HeLa00}; \cite{MMV}), the observed ratio of blue to red supergiants in the SMC
\cite{MMVII}, the variations with the metallicity of the number ratio of WR to O-type stars and of type Ibc to type II supernovae (\cite{MMX}; \cite{MMXI}). Here we apply the same physics for computing PopIII and very metal poor stars.

\begin{figure}
  \includegraphics[height=.3\textheight]{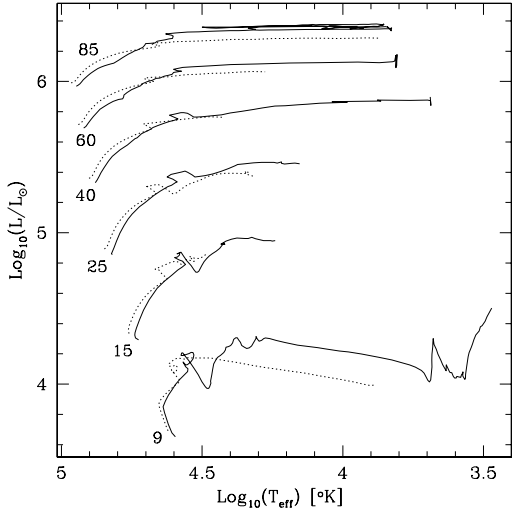}
  \caption{Evolution of $Z=0$ models in the Hertzsprung Russel diagram (Ekstr\"om et al. in preparation). The models with rotation are plotted with continuous lines, and those without rotation with dotted lines.}
         \label{fdhr}
\end{figure}

\section{Rotating PopIII stellar models}

Evolutionary tracks of non-rotating and rotating PopIII stellar models are shown in Fig.~\ref{fdhr}. The models are computed until the end of the core Si-burning, except the 9 M$_\odot$ that has developed a degenerate core before carbon ignition and has thus been stopped then, and the 15 M$_\odot$ model that has been stopped at the end of O-burning also because of a too degenerate core at that time. We chose an initial velocity of 800 km s$^{-1}$ on the ZAMS.
For the 60 M$_\odot$ model, this corresponds to a value of $\upsilon/\upsilon_{\rm crit}=0.52$ on the ZAMS, which is slightly superior to 0.4, the value required at solar metallicity to obtain averaged velocities during the MS phase corresponding to observed values. 

All the models, except the 9 M$_\odot$, reach the critical velocity during the MS phase. The mass which is lost by the mechanical winds amounts only to a few percents of the initial stellar mass and thus does not
much affect neither their evolution, nor their nucleosynthetic outputs. Much more mass can be lost by mechanical winds when the effects of magnetic fields are accounted for as prescribed in the Tayler-Spruit dynamo theory (\cite{Spruit02}, see for instance the paper by Ekstr\"om in these proceedings), or when the metallicity is non-zero. To illustrate this last point let us mention that non-rotating 60 M$_\odot$ stellar models at $Z$ equal to 0.00001 and 0.00050 lose during the MS phase respectively 0.21 and 0.78 M$_\odot$.
When $\upsilon_{\rm ini}=800$ km s$^{-1}$, mechanical winds bring away respectively 6.15 and 20.96 M$_\odot$!
Let us note also that the mass lost is enriched in H-burning products. This fact has been used by \cite{DecressinI} to explain the origin of the O-poor stars in galactic globular clusters.

\begin{figure}
  \includegraphics[height=.3\textheight]{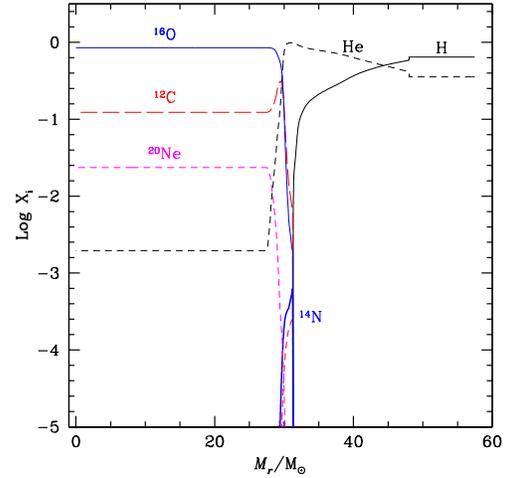}
  \caption{Variation of the abundances in mass fraction of some elements as a function of the Lagrangian mass in solar units, inside a PopIII ($Z=0$) 60 M$_\odot$ stellar model at the end of the core He-burning phase. The velocity on the ZAMS is 52\% of the critical velocity.}
         \label{chem0}
\end{figure}

\begin{figure}
  \includegraphics[height=.3\textheight]{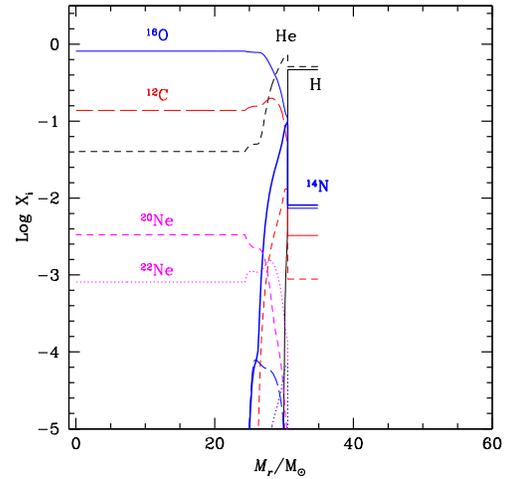}
  \caption{Same as Fig.~\ref{chem0} for a $Z=10^{-5}$ 60 M$_\odot$ stellar model. The velocity on the ZAMS is 65\% of the critical velocity.}
         \label{chem5}
\end{figure}

More striking are the effects of rotational mixing. As shown by \cite{EksNICIX}, the quantities of primary nitrogen produced by PopIII rotating models far exceed those obtained in non-rotating ones. Typically, the 60 M$_\odot$ non-rotating model ejects a few 10$^{-7}$ M$_\odot$ of primary nitrogen, while the model with $\upsilon_{\rm ini}=800$ km s$^{-1}$ ejects a few 10$^{-3}$ M$_\odot$ of this element, {\it i.e.} 4 orders of magnitude more. This has important impact on the chemical evolution of metal poor environment (see the paper
by Chiappini in the present volume).
It has to be emphasized here that non-rotating models can also produce some large amounts of primary nitrogen, however the mass and metallicity ranges where this occurs are in general quite limited. The process we invoke here for primary nitrogen production is operating not only on the whole mass range of stars
(including intermediate mass stars) but also on a large range of metallicities. Only when the metallicity becomes higher than about 0.001, does this mechanism no longer work. This is due to the fact that only at sufficiently low metallicity, is rotational mixing efficient enough to transport carbon and oxygen from the He-burning zones into the H-burning ones.

\begin{figure}
  \includegraphics[height=.3\textheight]{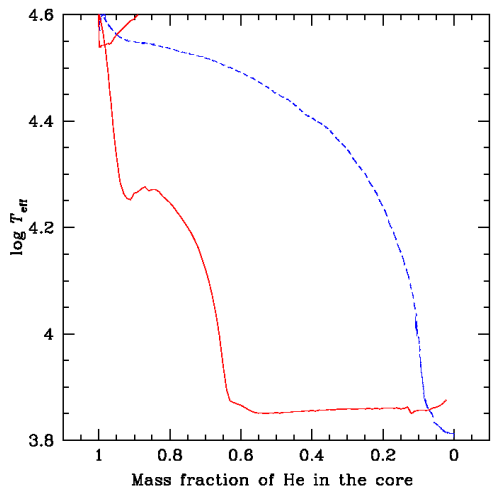}
  \caption{Evolution of the effective temperature during the core He-burning phase for a PopIII (dashed line) and
  a $Z=10^{-5}$ (continuous line) 60 M$_\odot$ stellar model.}
         \label{teff}
\end{figure}

Although the production of primary nitrogen is already quite significant in the rotating PopIII stars, very metal poor stars are still more efficient producers. This can be seen by comparing Figs~\ref{chem0} and \ref{chem5}. We see that the bump of primary nitrogen is much more pronounced in the $Z=10^{-5}$ model than in the PopIII model for the same initial rotation. This somewhat precises what we said above, namely that the efficiency of mixing (and thus the primary nitrogen production) increases when the metallicity decreases. Indeed, something special occurs when $Z=0$: massive PopIII stars begin the core H-burning phase by transforming hydrogen into helium through the pp chains since the CNO elements are absent. The pp chains produce energy at a too slow rate to be able to compensate for the energy lost at the surface, therefore the star has to extract energy from its gravitational reservoir and contracts. The slow contraction stops when the central temperature becomes high enough for triple alpha reactions to be activated. The $3\alpha$ reactions produce some carbon (of the order of 10$^{-10}$ in mass fraction) which activates then the usual CNO cycle. From this stage on, hydrogen burns in the core as in more metal rich stars {\it i.e.} through the CNO cycle. 
There is however a difference, it burns at a much higher temperature, at a temperature very near the one required
to burn helium. This means that at the end of the MS phase, the core does not need to contract a lot in order to reach the temperatures required for helium-burning. As a consequence the star does not evolve into the red part of the HR diagram as can be seen in Fig.~\ref{teff}. Due to this fact, the star undergoes much less mixing during the core He-burning phase. Indeed, no redwards evolution means that the gradient of the angular velocity at the border of the He-burning core remains much shallower and triggers much less mixing. 

When a small amount of metals is present, stars begin to burn their hydrogen from the start through the CNO cycle. Thus the central temperatures during the MS phases are well below those required for He-burning.
At the end of the MS phase, the core contracts and the star evolves to the red.
A steep angular velocity gradient appears at the core border and triggers strong mixing. 

Interestingly another consequence results from this rapid evolution to the red of rotating very metal poor stars: a convective zone appears at the surface which dredges-up at the surface great quantities of CNO elements, leading to the loss of a very large amounts of mass.
In our 60 M$_\odot$ stellar model with $Z=10^{-8}$ and $\upsilon_{\rm ini}=800$ km s$^{-1}$, the CNO
content at the surface amounts to one million times the one the star had at its birth \cite{Meynetal06}. Therefore the global metallicity at the surface becomes equivalent to that of a LMC stars while the star began its life with a metallicity which was about 600 000 times lower! If we apply the same rules used at higher metallicity relating the mass loss rate to the global metallicity, we obtain that the star may lose about half
of its initial mass due to this effect\footnote{Note that at the moment it is not possible to know if such a rule would apply in those circumstances, {\it i.e.} for typically a 60 M$_\odot$ PopIII star with an effective temperature of about 6000 K and a luminosity $\lg L/L_\odot=6.1$.}. As shown by \cite{Meynetal06} and \cite{Hir07} the matter released
by these winds is enriched in both H- and He-burning products and present striking similarities with the abundance patterns observed at the surface of C-rich Ultra-Metal-Poor-Stars (CRUMPS). This process does not occur in the present PopIII stellar models. As can be deduced from the explanations above, PopIII stars undergo less efficient mixing during the core He-burning phase. Thus the H-burning shell and the outer radiative envelope are enriched at a lower rate in CNO elements. This tends to delay the apparition of an outer convective zone and to reduce its extension. The increase of the surface metallicity remains very modest and does not lead to strong mass loss.

From the above results, we see that rotational mixing appears as the dominant effect of rotation for PopIII stars, modifying significantly the amount of primary nitrogen. The presence of a little amount of metals suffices to boost the efficiency of rotational mixing and the importance of mass loss. In that respect metallicity is like
the salt of the cosmos: a small amount is sufficient to enhance its flavor!

\section{``Spinstars'' at very low metallicities?} 

Let us call ``spinstars'' those stars with a sufficiently high initial rotation in order to have their evolution significantly affected by rotation. In this section, we present some arguments supporting the view according to which spinstars might have been more common in the first generations of stars in the Universe.
A direct way to test this hypothesis would be to obtain measures of surface velocity of very metal poor massive stars and to see whether their rotation is superior to those measured at solar metallicity. At the moment, such measures can be performed only for a narrow range of metallicities for $Z$ between 0.002
and 0.020. Interestingly already some effects can be seen. For instance \cite{Keller04} presents measurements of the projected rotational velocities of a sample of 100 early B-type main-sequence stars in the Large Magellanic Cloud (LMC). He obtains that the stars of the LMC are more rapid rotators than their Galactic counterparts and that, in both galaxies, the cluster population exhibits significantly more rapid rotation than that seen in the field (a point also recently obtained by \cite{HuangGies06}). More recently \cite{Martayan07} obtain that the angular velocities of B (and Be stars) are higher in the SMC than in the LMC and MW. 

For B-type stars, the higher values obtained at lower $Z$ can be the result of two processes: 1) the process of star formation produces more rapid rotators at low metallicity; 2) the mass loss
being weaker at low $Z$, less angular momentum is removed from the surface and thus starting from the same initial velocity, the low $Z$ star would be less slowed down by the winds. In the case of B-type stars, the mass loss rates are however quite modest and we incline to favor the first hypothesis, {\it i.e.} a greater fraction of fast rotators at birth at low metallicity. Another piece of argument supporting this view is the following: in case the mass loss rates are weak (which is the case on the MS phase for B-type stars), then the surface velocity is mainly determined by two processes, the initial value on the ZAMS and the efficiency of the angular momentum transport from the core to the envelope. In case of very efficient transport, the surface will receive significant amount of angular momentum transported from the core to the envelope. The main mechanism responsible for the transport of the angular momentum is meridional circulation. The velocities of the meridional currents in the outer layers are smaller when the density is higher thus in more metal poor stars. Therefore, starting from the same initial velocity on the ZAMS, one would expect that B-type stars at solar metallicity (with weak mass loss) would have higher surface velocities than the corresponding stars at low $Z$. The opposite trend is observed. Thus, in order to account for the higher velocities of B-type stars in the SMC and LMC, in the frame of the present rotating stellar models, one has to suppose that stars on the ZAMS have higher velocities at low $Z$. Very interestingly, the fraction of Be stars (stars rotating near the critical velocity) with respect to the total number of B stars is higher at low metallicity (\cite{MaederGrebel99}; \cite{Wisniewski06}). This confirms the trend
discussed above favoring a  higher fraction of fast rotators at low $Z$.

There are at least four other striking observational facts which might receive an explanation based on massive fast rotating models. {\it First}, indirect observations indicate the presence of very helium-rich stars in the globular cluster $\omega$Cen \cite{Piotto05}. Stars with a mass fraction of helium, $Y$,  equal to 0.4 seem to exist, together with a population of normal helium stars with $Y=0.25$. 
Other globular clusters appear to host helium-rich stars \cite{caloi07}, thus
the case of $\omega$Cen is the most spectacular but not the only one.
There is no way for these very low mass stars to enrich their surface in such large amounts of helium and thus they must have formed from protostellar cloud having such a high amount of helium. Where does this helium come from? We proposed that it was shed away by the winds of metal poor fast rotating stars \cite{MMocen}. 

{\it Second}, in globular clusters, stars made of material only enriched in H-burning products have been observed (see the review by \cite{Gratton04}). Probably these stars are also enriched in helium and thus this observation is related to the one reported just above. The difference is that proper abundance studies can be performed for carbon, nitrogen, oxygen, sodium, magnesium, lithium, fluorine \dots, while for helium only indirect inferences based on the photometry can be made. \cite{DecressinI}  propose that the matter from which the stars rich in H-burning products are formed, has been released by slow winds of fast rotating massive stars. Of course, part of the needed material can also be released by AGB stars. The massive star origin presents however some advantages: first a massive star can induce star formation in its surrounding, thus two effects, the enrichment and the star formation can be triggered by the same cause. Second, the massive star scenario allows to use a less flat IMF than the scenario invoking AGB stars \cite{PC06}. The slope of the IMF might be even a Salpeter's one in case the globular cluster lost a great part of its first generation stars by tidal stripping (see Decressin et al. in preparation).

{\it Third}, the recent observations of the surface abundances of very metal poor halo stars\footnote{These stars are in the field and present [Fe/H] as low as -4, thus well below the metallicities of the globular clusters.} show the need of a very efficient mechanism for the production of primary nitrogen \cite{Chiappinial05}. As explained in the paper by Chiappini in the present proceedings (see also \cite{Chiappinial06}), a very nice way to explain this very efficient primary nitrogen production is to invoke fast rotating massive stars. Very interestingly, fast rotating massive stars help not only in explaining the behavior of the N/O ratio at low metallicity but also those of the C/O and of $^{12}$C/$^{13}$C ratios.

{\it Fourth}, below about [Fe/H] $<$ -2.5, a significant fraction of very iron-poor stars are C-rich
(see the review by \cite{BeersCrist05}). Some of these stars show no evidence of $s$-process enrichments by AGB stars and are thus likely formed from the ejecta of massive stars. The problem is how to explain the very high abundances with respect to iron of CNO elements.
\cite{Meynetal06} and \cite{Hir07} proposed that these stars might be formed from the winds of very metal poor fast rotating stars. It is likely that rotation also affects the composition of the ejecta of intermediate mass stars. \cite{Meynetal06} predict the chemical composition of the envelope of a 7 M$_\odot$ E-AGB star which have been enriched by rotational mixing. The composition presents striking similarities with the abundance patterns observed at the surface of CRUMPS. The presence of overabundances of fluorine (see the paper by Schuler in the present volume) and of $s$-process elements might be used to discriminate between massive and intermediate mass stars.

All the above observations seem to point towards the same direction, an important population of spinstars at low Z. How many? What is the origin of the fast rotation? What are the consequences for the Gamma ray Burst progenitors?
All these questions have still to be addressed in a quantitative way and offer nice perspective for future works. 




\bibliographystyle{aipproc}   
\def\aj{AJ}%
\def\actaa{Acta Astron.}%
\def\araa{ARA\&A}%
\def\apj{ApJ}%
\def\apjl{ApJ}%
\def\apjs{ApJS}%
\def\ao{Appl.~Opt.}%
\def\apss{Ap\&SS}%
\def\aap{A\&A}%
\def\aapr{A\&A~Rev.}%
\def\aaps{A\&AS}%
\def\azh{AZh}%
\def\baas{BAAS}%
\def\bac{Bull.~astr.~Inst.~Czechosl.}%
\def\caa{Chinese Astron. Astrophys.}%
\def\cjaa{Chinese J. Astron. Astrophys.}%
\def\icarus{Icarus}%
\def\jcap{J. Cosmology Astropart. Phys.}%
\def\jrasc{JRASC}%
\def\mnras{MNRAS}%
\def\memras{MmRAS}%
\def\na{New A}%
\def\nar{New A Rev.}%
\def\pasa{PASA}%
\def\pra{Phys.~Rev.~A}%
\def\prb{Phys.~Rev.~B}%
\def\prc{Phys.~Rev.~C}%
\def\prd{Phys.~Rev.~D}%
\def\pre{Phys.~Rev.~E}%
\def\prl{Phys.~Rev.~Lett.}%
\def\pasp{PASP}%
\def\pasj{PASJ}%
\def\qjras{QJRAS}%
\def\rmxaa{Rev. Mexicana Astron. Astrofis.}%
\def\skytel{S\&T}%
\def\solphys{Sol.~Phys.}%
\def\sovast{Soviet~Ast.}%
\def\ssr{Space~Sci.~Rev.}%
\def\zap{ZAp}%
\def\nat{Nature}%
\def\iaucirc{IAU~Circ.}%
\def\aplett{Astrophys.~Lett.}%
\def\apspr{Astrophys.~Space~Phys.~Res.}%
\def\bain{Bull.~Astron.~Inst.~Netherlands}%
\def\fcp{Fund.~Cosmic~Phys.}%
\def\gca{Geochim.~Cosmochim.~Acta}%
\def\grl{Geophys.~Res.~Lett.}%
\def\jcp{J.~Chem.~Phys.}%
\def\jgr{J.~Geophys.~Res.}%
\def\jqsrt{J.~Quant.~Spec.~Radiat.~Transf.}%
\def\memsai{Mem.~Soc.~Astron.~Italiana}%
\def\nphysa{Nucl.~Phys.~A}%
\def\physrep{Phys.~Rep.}%
\def\physscr{Phys.~Scr}%
\def\planss{Planet.~Space~Sci.}%
\def\procspie{Proc.~SPIE}%

\bibliography{bibADS}


\end{document}